\documentclass[%
 reprint,
superscriptaddress,
 amsmath,amssymb,
 aps,
prb,
]{revtex4-2}
\usepackage{graphicx}
\usepackage{dcolumn}
\usepackage{bm}
\usepackage{hyperref}
\usepackage{nicematrix}
\usepackage{float}
\begin{document}


\title{First-principles exploration of superconductivity in intercalated bilayer borophene phases}

\author{Božidar N. Šoškić}
\affiliation{%
 Faculty of Natural Sciences and Mathematics, University of Montenegro, Džordža Vašingtona bb, 81000 Podgorica, Montenegro}%
 \affiliation{Department of Physics \& NANOlab Center of Excellence, University of Antwerp, Groenenborgerlaan 171, B-2020 Antwerp, Belgium}

\author{Jonas Bekaert}
\email{jonas.bekaert@uantwerpen.be}
\affiliation{Department of Physics \& NANOlab Center of Excellence, University of Antwerp, Groenenborgerlaan 171, B-2020 Antwerp, Belgium}

\author{Cem Sevik}
\affiliation{Department of Physics \& NANOlab Center of Excellence, University of Antwerp, Groenenborgerlaan 171, B-2020 Antwerp, Belgium}
\affiliation{
 Department of Mechanical Engineering, Faculty of Engineering, Eskisehir Technical University, 26555 Eskisehir, Turkey
}

\author{Željko Šljivančanin}
\affiliation{Vinča Institute of Nuclear Sciences, National Institute of the Republic of Serbia, University of Belgrade, P. O. Box 522, RS-11001 Belgrade, Serbia}

\author{Milorad V. Milošević}
\email{milorad.milosevic@uantwerpen.be}
\affiliation{Department of Physics \& NANOlab Center of Excellence, University of Antwerp, Groenenborgerlaan 171, B-2020 Antwerp, Belgium}

\date{\today}

\begin{abstract}
We explore the emergence of phonon-mediated superconductivity in bilayer borophenes by controlled intercalation with elements from the groups of alkali, alkaline-earth, and transition metals, using systematic first-principles and Eliashberg calculations. We show that the superconducting properties are primarily governed by the interplay between the out-of-plane ($p_{z}$) boron states and the partially occupied in-plane ($s+p_{x,y}$) bonding states at the Fermi level. Our Eliashberg calculations indicate that intercalation with alkaline-earth elements leads to the highest superconducting critical temperatures ($T_{c}$). Specifically, Be in $\delta_{4}$, Mg in $\chi_{3}$, and Ca in the kagome bilayer borophene demonstrate superior performance with $T_{c}$ reaching up to 58~K. Our study therefore reveals that intercalated bilayer borophene phases are not only more resilient to chemical deterioration, but also harbor enhanced $T_{c}$ values compared to their monolayer counterparts, underscoring their substantial potential for the development of boron-based two-dimensional superconductors.

\end{abstract}

\maketitle


\section{Introduction}

Possessing a rich research history, elemental boron continues to captivate the scientific community with its exceptional and distinctive chemical properties~\cite{his1, his2}. In the first phase of boron exploration, the primary objective was to comprehend the fundamental properties of this lightweight element by scrutinizing its behavior across different dimensionalities, mainly encompassing three-dimensional (3D) clusters, and the prospect of two-dimensional (2D) structures~\cite{boustani,liu,2d_3d}. Building upon prior theoretical predictions~\cite{tp1,tp2}, a monumental leap forward was accomplished in 2015 when 2D boron structures were synthesized on a silver substrate, and named borophenes~\cite{exp1,exp2}. These borophenes exhibit the same structural planarity as graphene, but also an abundance of hexagonal voids, with different concentrations (quantified by the $\eta$ parameter) and configurations~\cite{voids}. Borophenes are highly versatile and can be synthesized on an extensive range of metallic substrates, showcasing their adaptability and potential for widespread applications~\cite{podloga1,podloga2,podloga3}. This major breakthrough has facilitated numerous research milestones, demonstrating the distinctive physical properties of borophene, such as high mechanical flexibility and strength~\cite{m1,m2}, optical transparency~\cite{opt}, the presence of Dirac fermions~\cite{dirac1,dirac2}, and theoretically predicted phonon-mediated superconductivity~\cite{sup1,sup2}. This makes borophene highly suitable for a plethora of technological applications, such as energy storage~\cite{energy}, gas sensing~\cite{gas1,gas2}, catalysis~\cite{cat}, and versatile superconducting devices. Regarding the latter, borophenes have been theoretically predicted to exhibit critical temperatures ($T_{c}$) of up to 33 K, and a potential for two-gap superconductivity ~\cite{gao,multigap,kagome1}. However, in realistic conditions, the presence of the substrate may significantly impact these predictions. For example, simulations incorporating the effects of strain and electron doping revealed a notable suppression of superconductivity in the $\beta_{12}$ phase ($\eta=1/6$)~\cite{suppressed}.

\begin{figure*}[!htp]
    \includegraphics[width=\textwidth]{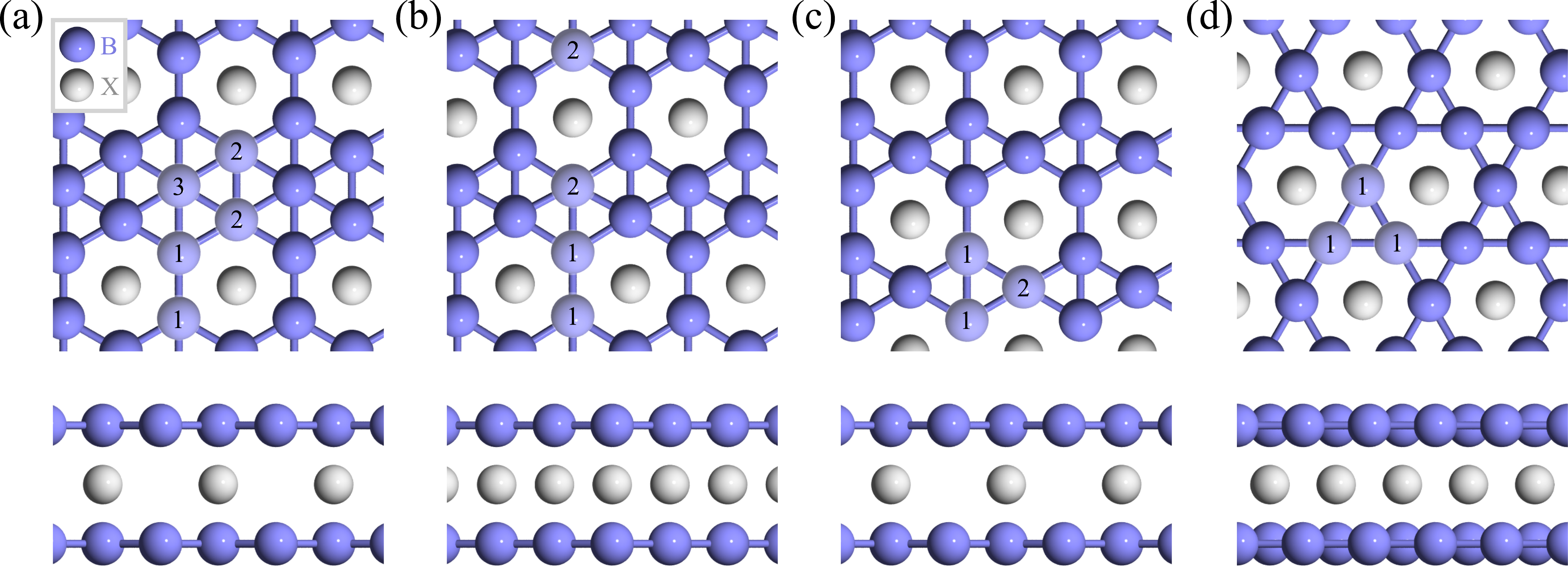}
\caption{\label{fig:structures} Top and side view of intercalated (a) $\beta_{12}$, (b) $\chi_{3}$, (c) $\delta_{4}$, and (d) kagome bilayer borophene phases. Boron and intercalated atoms are represented by blue and grey spheres, respectively.}
\end{figure*}

Borophene's susceptibility to oxidation in ambient environments forms one of the main hindrances to its controlled functional properties~\cite{ox1,ox2}. One approach to remedy this issue is surface functionalization~\cite{func}, however, an alternative research pathway has emerged with the recent synthesis of bilayer borophene. Both zigzag $\beta_{12}$-like borophene on a Cu(111) substrate~\cite{bl1} and $\alpha$ phase ($\eta=1/9$) on a Ag(111) substrate have been synthesized in bilayer form~\cite{bl2}. These bilayer structures exhibit exceptional thermal stability, and resilience to oxidation~\cite{bl3}, and hold promise for optical sensors and thermoelectric devices~\cite{bl5}, as well as anode materials for future Li-ion batteries~\cite{batteries}. Theoretical examinations have shown that certain bilayer configurations display covalent interlayer bonding phenomena that not only influence the localization of electronic states but also induce partial decoupling of the entire bilayer from the substrate, ultimately leading to a ``quasi'' free-standing system~\cite{quasi}. Hence, these double-layered configurations provide sufficient means to mitigate surface reactivity of borophenes as well as adverse substrate effects. Moreover, through liquid-phase exfoliation it is possible to fabricate freestanding monolayer~\cite{freestanding1}, as well as bilayer borophenes~\cite{freestanding2}.  

These advances have opened up an entire exploration avenue, specifically in the quest for novel functionalities in boron-based 2D materials. Particulary, bilayer structures offer convenient intercalation opportunities to tailor desired properties with relative ease. Theoretical investigations have yielded support for the viability of the latter, as exemplified by the predicted superconductivity of Mg-intercalated $\beta_{12}$, $\chi_{3}$ ($\eta=1/5$), $\delta_{4}$ and kagome (both with $\eta=1/4$) bilayer borophene with $T_{c}$ of up to 13~K~\cite{mgb6,mg_intercalated}. Moreover, Ca-intercalated kagome bilayer borophene has been demonstrated to have a $T_{c}$ of 23~K~\cite{PRM} and 36~K in the MBene ($\eta=1/3$) form~\cite{cahex}. Furthermore, when bilayer borophenes are stacked, their 3D der Waals structures~\cite{3d} could also prove a fertile ground for phonon-mediated superconductivity~\cite{sup_sta}.

Building upon these recent advances in the field, the primary aim of this manuscript is to illuminate pathways to optimized phonon-mediated superconductivity in $\beta_{12}$, $\chi_{3}$, $\delta_{4}$, and kagome bilayer borophene through strategic intercalation with specific elements from the group of alkali, alkaline earth, and transition metals. The main goal is to understand the mechanisms promoting stability and robust 2D superconductivity in these systems. The manuscript is organized as follows. Sec.\ II contains the methodology used in this work. In Sec.\ III we present the results on structural, electronic, vibrational, and superconducting properties of the studied structures. In Sec.\ IV we present the results of fully anisotropic Migdal-Eliashberg calculations for the most prominent candidates for superconductivity (as identified on the isotropic level). Finally, Sec.\ V summarizes our findings and conclusions.
\begin{table*}[!htp]
\caption{\label{tab_a} Space group, $\eta$ value, number of B atoms in the unit cell ($N_{B}$), number of B atoms ($N_{at}$) with CN = 4, 5, 6, lattice parameters, and the average interlayer distance $\langle d\rangle$, for different bilayer borophene structures studied in this work.}
\begin{ruledtabular}
\begin{tabular}{cccccccc}
Structure & Space group & $\eta$ &$N_{B}$ & $N_{at}$ with CN = 4, 5, 6& $a$~(\AA) & $b$~(\AA) & $\langle d\rangle$~(\AA)\\\hline
$\beta_{12}$ & $Pmm2$ (no. 25) & $1/6$ &10& 4:4:2&2.943 &5.073& 2.761\\
$\chi_{3}$& $Cmmm$ (no. 65) &$1/5$& 8&4:4:0&4.451 & - &3.166\\
$\delta_{4}$& $Pmmm$ (no. 47)& $1/4$&6&6:0:0 &2.889 & 3.332 & 4.957\\
Kagome& $P6/mmm$ (no.
191) & $1/4$ &6&6:0:0&3.316 & - & 4.830\\ 
\end{tabular}
\end{ruledtabular}
\end{table*}
\section{Methodology}

The structural and electronic properties were examined via Density Functional Theory (DFT), as implemented within the Quantum ESPRESSO package~\cite{qe1,qe2}, using the Perdew-Burke-Ernzerhof (PBE) functional and norm-conserving pseudopotentials from the PseudoDojo project~\cite{pseudodojo}. The wave functions and the electron density were expanded in plane waves with cutoff energies of 120~Ry and 480~Ry, respectively. A vacuum region of 20~{\AA} was applied to prevent the interaction of periodic images. For the $\beta_{12}$ phase a 24$\times$18$\times$1 Monkhorst-Pack electronic wave-vector ($\mathbf{k}$) grid was chosen, and a 24$\times$24$\times$1 grid was used for the integration of electronic states over the $\mathbf{k}$-space for the $\chi_{3}$, $\delta_{4}$ and kagome phases. All structures underwent a complete relaxation so that the Hellmann-Feynman forces on each atom were reduced to below $10^{-5}$~Ry/bohr. Generally, a Methfessel-Paxton smearing parameter of 0.02~Ry was selected to treat the electronic occupation around the Fermi level. For certain compounds this smearing value results in soft phonon branches, suggestive of potential unstable phonon modes. Therefore, to validate the structural stability, phonon dispersions were also calculated at smaller smearing values of 0.01, 0.005, and 0.0025 Ry. We find that at lower smearing values pristine $\delta_4$ phase becomes dynamical unstable. Subsequently, the dynamical properties of the system were examined at the harmonic level, using the Density Functional Perturbation Theory (DFPT) method~\cite{dfpt}, using 8$\times$6$\times$1, 8$\times$8$\times$1, 8$\times$8$\times$1, and 12$\times$12$\times$1 $\mathbf{q}$-grids for $\beta_{12}$, $\chi_{3}$, $\delta_{4}$, and kagome bilayer borophene, respectively. To characterize the superconducting state we employed the isotropic Eliashberg theory -- a quantitative and precise expansion upon the Bardeen-Cooper-Schrieffer (BCS) theory~\cite{eliashberg_1,eliashberg_2}. Using the density of states at the Fermi level ($N(0)$), the matrix elements of the electron-phonon coupling ($g_{\mathbf{k},\mathbf{k+q},\nu}$), the phonon ($\omega_{\mathbf{q},\nu}$) and electron ($\epsilon_{\mathbf{k}}$) bands obtained from our \textit{ab initio} calculations, we have evaluated the isotropic Eliashberg spectral function using the following expression: 
\begin{equation}
    \alpha^{2}F=\frac{1}{N(0)}\sum_{\mathbf{k},\mathbf{q}}\sum_{\nu}|g_{\mathbf{k},\mathbf{k+q},\nu}|^{2}\delta(\epsilon_{\mathbf{k+q}})\delta(\epsilon_{\mathbf{k}})\delta(\omega-\omega_{\mathbf{q},\nu}).
\end{equation}
Further, the electron-phonon coupling (EPC) constant was calculated as:
\begin{equation}
\lambda=2\int_{0}^{\infty}\frac{\alpha^{2}F(\omega)}{\omega}d{\omega}.
\end{equation}
Based on this, we reported the $T_{c}$ values using the McMillan-Allen-Dynes formula~\cite{allen_1, allen_2}, with a value of 0.1 for the Morel-Anderson pseudopotential describing the Coulomb interaction within the Cooper pairs ($\mu^*$).  

For materials predicted to possess high $T_{c}$ values, self-consistent solutions of the fully anisotropic Migdal-Eliashberg equation were obtained using the Electron-Phonon Wannier (EPW) code~\cite{epw1, epw2}. For the initial guesses of Maximally Localized Wannier Functions (MLWFs), we used $s$-, $p_z$-, $p_x$-, and $p_y$-like orbitals for B atoms, and $s$-like orbitals for intercalated atoms. The computation of electron-phonon matrix elements was first executed on coarse $\mathbf{k}$- and $\mathbf{q}$-grid -- the same as in the isotropic step. These were subsequently interpolated onto a finer (denser) 192$\times$192$\times$1, 216$\times$216$\times$1, and 216$\times$216$\times$1 $\mathbf{k}$-, and 64$\times$64$\times$1, 72$\times$72$\times$1, and 108$\times$108$\times$1 $\mathbf{q}$-grid for Mg in $\chi_3$, Be in $\delta_4$, and Ca in kagome bilayer borophene, respectively, to accurately obtain the electron-phonon coupling constants ($\lambda$) and superconducting gaps. The
cutoff for the fermionic Matsubara frequencies $\omega_{j}=(2j+1)\pi T$ was set to 0.6~eV and $\mu^{*}$ was the same as in the isotropic calculations (0.1).

\section{Results and discussion}

\subsection{Crystal structures and stability}

The borophene phases consist of a sheet of boron (B) atoms arranged in a triangular lattice configuration, exhibiting a distinct pattern of empty close-packed hexagonal holes (HH), arising from the absence of B atoms at their center ($B_{hollow}$). The remaining $B_{hollow}$ atoms play the role of electron donors, while the HH themselves serve as electron acceptors, engendering an atypical chemical mechanism of ``self-doping''. In order to classify different phases of borophene, it is of great importance to consider the coordination number (CN) of the B atoms. As depicted in Table~\ref{tab_a}, the four borophene phases under consideration exhibit variations in both the number of atoms within the unit cell and their corresponding CN values. Fig.~\ref{fig:structures} illustrates that while the $\delta_{4}$ and kagome phases share the same CN value for their atoms, they differ in their lattice symmetries. Specifically, the former comprises two types of symmetrically arranged atoms (labeled as 1 and 2), whereas the latter consists of only one type.

\begin{figure}[!hbp]
    \includegraphics[width=8.6 cm]{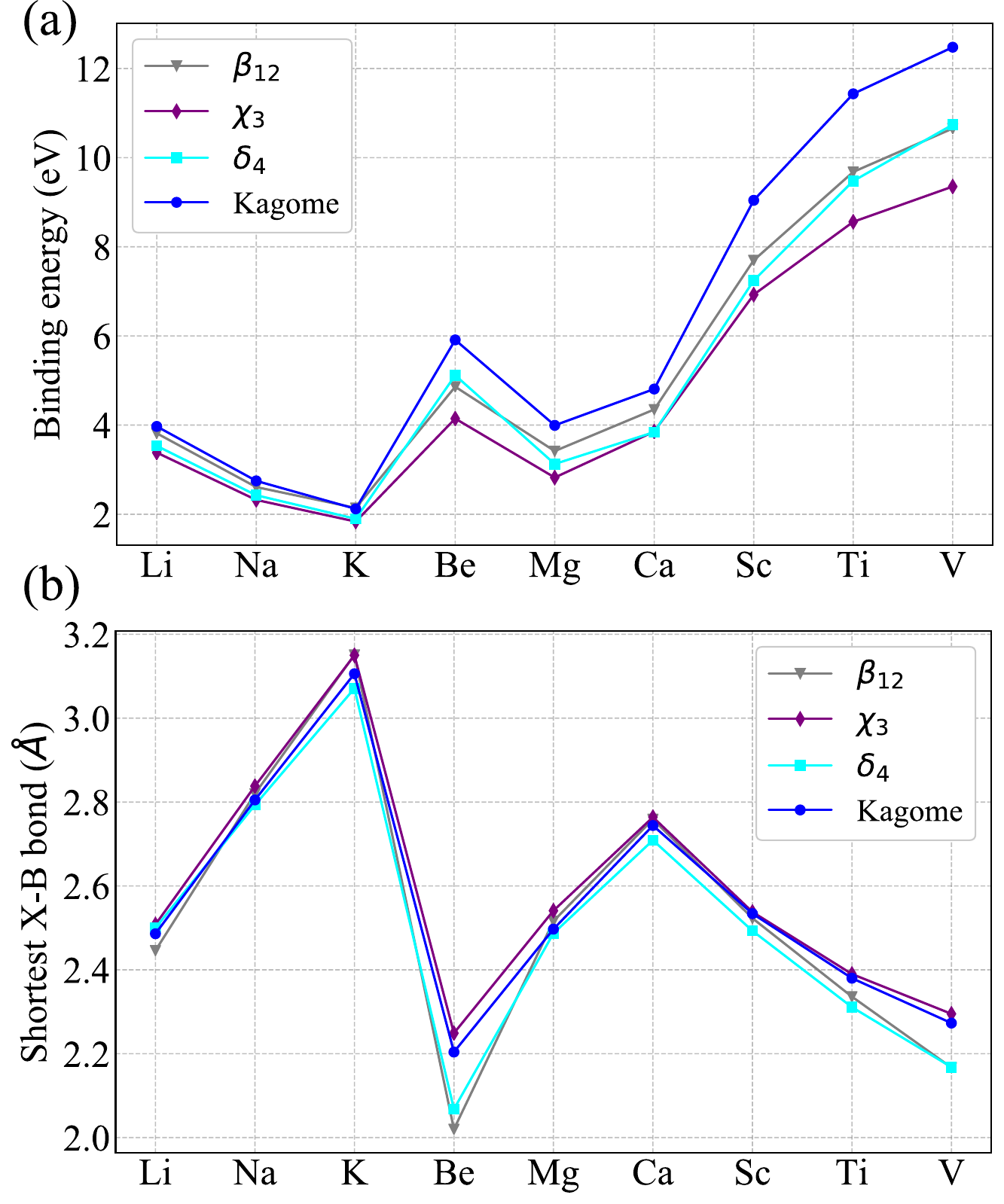}
\caption{\label{fig:energies} (a) Binding energy of intercalated bilayer borophene phases, together with (b) the corresponding minimal X-B distances.}
\end{figure}

The CN values also play a significant role in driving variations in the local electronic properties of the B atoms, with a direct effect on their reactivity. Here, we focused on bilayer borophene and its properties upon intercalation with selected metals. To characterize different adsorption geometries of intercalants, we calculated the binding energy,
\begin{equation}
    E_{b}(X)=\frac{N_{X}E_{X}+E_{BB}-E_{BXB}}{N_{X}},
\end{equation}
where $N_{X}$ is the number of X atoms in the unit cell (in the present work $N_{X}=1$ in all cases), and $E_{X}$, $E_{BB}$, $E_{BXB}$ are the total energies of the intercalants in isolated form, and of pristine and intercalated bilayer borophenes, respectively. 
We found a strong preference for metal atoms to bind between vertically aligned HH sites of the bilayers with the AA stacking of borophene sheets, as shown in Fig.~\ref{fig:structures}. The set of examined binding sites and the corresponding binding energies is presented in the Supplementary Material (SM), Fig.\ S2~\cite{noteSI}.

Distinct trends can thus be distilled for different metal intercalants. For instance, the binding energy of the K atom exhibits the lowest values in all phases of bilayer borophene [see Fig.~\ref{fig:energies}(a)].  This is in line with its minimal electronegativity value, as follows from the Bader charge analysis, presented in the SM (Table S6)~\cite{noteSI}). Comparing different borophene phases, an overall trend in the binding energy emerges: strongest binding for the kagome phase, followed by $\beta_{12}$, $\delta_{4}$, and ultimately $\chi_{3}$. There are just a few exceptions to this rule, with minor energy differences between subsequent borophene phases. 

Furthermore, specific interactions between the intercalant and B atoms with different CN lead to variations in their bond lengths. In the $\beta_{12}$ phase, we observed that the shortest X-B bond occurs for B$_{1}$ atoms [see Fig.~\ref{fig:structures}], with its lowest value for Be (2.02~\AA) and highest for K (3.15~\AA), as depicted in Fig.~\ref{fig:energies}(b). In the case of $\chi_{3}$ the shortest X-B bond is consistently formed with the B$_2$ atoms. Finally, the $\delta_{4}$ structure is characterized by minimal bonds with either B$_{1}$ or B$_{2}$, depending on the type of intercalant (B$_{1}$ for alkali, and B$_{2}$ for alkaline-earth and transition metals). Fig.\ \ref{fig:energies} shows that the binding energies anticorrelate with the X-B bond lengths. Therefore, in general, stronger bonding of the intercalants goes hand in hand with shorter bond lengths with the adjacent boron atoms, and with Ca as the only intercalant forming a minor exception to this rule. 

\begin{figure}[!h]
    \includegraphics[width=8.6 cm]{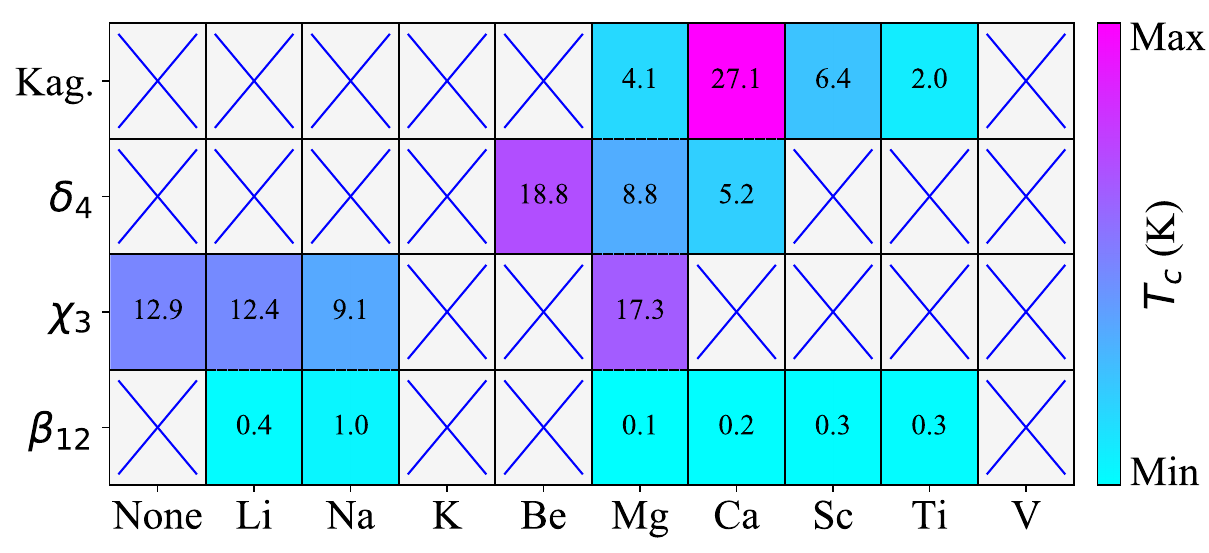}
\caption{\label{fig:tc} Overview of the dynamically stable bilayer borophene structures in conjunction with their calculated superconducting critical temperature ($T_c$). The blue crosses indicate dynamically unstable structures.}
\end{figure}

Let us now turn to the dynamical stability, based on phonon dispersions calculated using DFPT [see the SM, Fig.\ S32--S36~\cite{noteSI} for the collected phonon dispersions]. It is known that in the $\beta_{12}$ monolayer configuration, electron deficiency is minimal, and that this phase is dynamically stable when a small amount of uniform tensile strain is applied~\cite{sup1}. In a similar vein, we find that its bilayer counterpart can be stabilized through the inclusion of intercalants. Another dynamically stable monolayer phase is $\chi_{3}$ -- at least upon applying tensile strain~\cite{sup1}. We found its pristine bilayer AA-stacked configuration to be dynamically stable, as well as its intercalated form with alkali metals (Li, Na) and Mg as intercalants. Proceeding with the other phases, it becomes evident that the electron deficit increases significantly, resulting in dynamical instability of both the $\delta_{4}$ and kagome monolayers. The stability in their double-layered configurations is achieved by introducing intercalants capable of effectively large transfer of electrons, like some alkaline-earth and transition metals. During the process of intercalation, notable alterations occur in both the in-plane lattice constants and the average interlayer distances (cf.\ SM, Table S2--S5~\cite{noteSI}). It is important to note that all intercalated structures considered in the present study were found to be non-magnetic through spin-polarized calculations we have preformed, fully in line with earlier reports in the literature~\cite{mag1,mag2}. Fig.~\ref{fig:tc} presents an overview of our results on the dynamical stability, together with the corresponding values of $T_{c}$ for the found stable structures. 



\subsection{Electronic, vibrational, and superconducting properties}
\subsubsection{\texorpdfstring{$\beta_{12}$}{beta12} phase of bilayer borophene \texorpdfstring{$(\eta=1/6)$}{n_16}}

As the next step towards understanding their superconducting properties, we here discuss the electronic structure of the bilayer borophenes under study. The intrinsic metallic nature of $\beta_{12}$ bilayer systems is evident from the orbital-resolved band structures and projected electronic density of states (PDOS), presented in the SM, Figs.\ S8 and S16~\cite{noteSI}. The dominant contributions to this metallic nature stems primarily from the B-$p_{z}$ orbitals. In addition, there are relatively minor contributions from the $s$ orbitals in the case of alkali and alkaline-earth intercalants, and more substantial contributions from the $d$ orbitals for the transition metal intercalants. 

\begin{figure*}[!htp]
    \includegraphics[width=\textwidth]{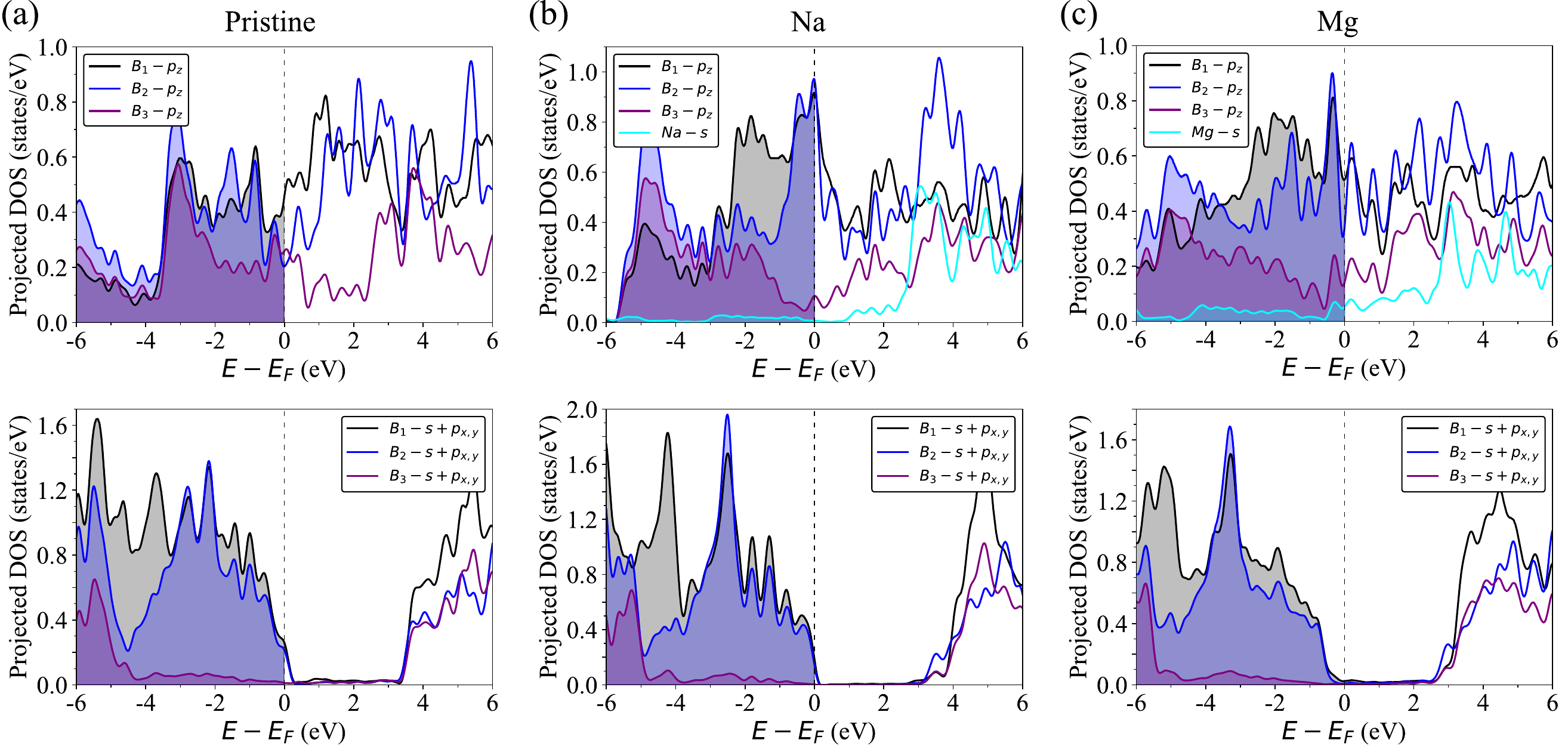}
\caption{\label{fig:dos_1} The DOS projected on the $p_{z}$ and $s+p_{x,y}$ orbitals of different B atoms (with the numbering of the atoms corresponding to the one in Fig.\ \ref{fig:structures}(a)) in (a) pristine, (b) Na, and (c) Mg intercalated $\beta_{12}$ bilayer borophene. In the case of the intercalated structures, $s$ orbital contributions from the X atoms are also shown.}
\end{figure*}
When compared to the other phases, the $\beta_{12}$ structure has the lowest $\eta$ value and the lowest electron deficit in the monolayer form. Fig.~\ref{fig:dos_1} clearly illustrates that upon intercalation the bonding states ($\sigma$) of the B in-plane orbitals ($s+p_{x,y}$) are almost fully occupied, exhibiting a gap of approximately 3~eV that separates them from their anti-bonding states ($\sigma^{*}$). The remaining electrons partially fill the states of the B out-of-plane orbitals ($p_{z}$). As expected, intercalation leads to a significant increase in the $p_{z}$-projected DOS at the Fermi level ($E_{F}$), stemming from B$_{1}$ and B$_{2}$ atoms. We found that Na intercalation leads to the highest total DOS at $E_{F}$, with a value of 2.33~states/eV, while Mg intercalation yields the lowest value, namely 1.08~states/eV. At $E_{F}$, the primary contribution to the total DOS for all structures arises from out-of-plane states. In addition, in the cases of Sc and Ti, $d$ orbitals also assume a more prominent role, contributing 0.64 and 0.68~states/eV, respectively. Furthermore, when compared to other dynamically stable structures, Na exhibits the highest DOS coming from not fully occupied $\sigma$ states at $E_{F}$, with a value of 0.33~states/eV [see Fig.\ \ref{fig:dos_1}(e)], followed by Sc and Ti with 0.19 and 0.16~states/eV, respectively.


As depicted in Fig.~\ref{fig:tc}, we found that the $\beta_{12}$ configuration is dynamically unstable in pristine form as well as upon intercalation with K, Be, and V. For all other compounds, the phonon dispersions clearly demonstrate dynamical stability, with several phonon modes having a significantly flat dispersion, producing pronounced peaks in the phonon DOS. Below the energy of $\sim$50~meV, the modes mainly originate from hybridized B and intercalant vibrations, while purely B-based phonon modes are situated above $\sim$50~meV. Our momentum-resolved electron-phonon coupling (EPC) analysis shows that the highest EPC values commonly correspond to softened phonon modes at $\sim$9~meV near the Y point [see SM, Fig.\ S37~\cite{noteSI}]. 

Now we proceed to investigate the superconducting properties based on isotropic Eliashberg theory. Overall, the total EPC values in these compounds are very low, reaching their maximum value of 0.35 in the case of Na, yielding $T_{c} \sim 1$ K. The two main peaks in the Eliashberg function [see SM, Figs.\ S37(b)~\cite{noteSI}] originate from the B-based ZA phonon mode and from the $A_{u}$ optical phonon mode (within the $C_{s}$ point group), corresponding to out-of-plane movement of both the Na and B atoms (at $\sim$30~meV). This stronger EPC in the case of Na compared to other intercalants correlates with its higher DOS at $E_{F}$. Nevertheless, in a broader context, the very modest $T_{c}$ of this phase renders it a less compelling candidate for further theoretical and experimental exploration.   

\begin{figure*}[!htp]
    \includegraphics[width=\textwidth]{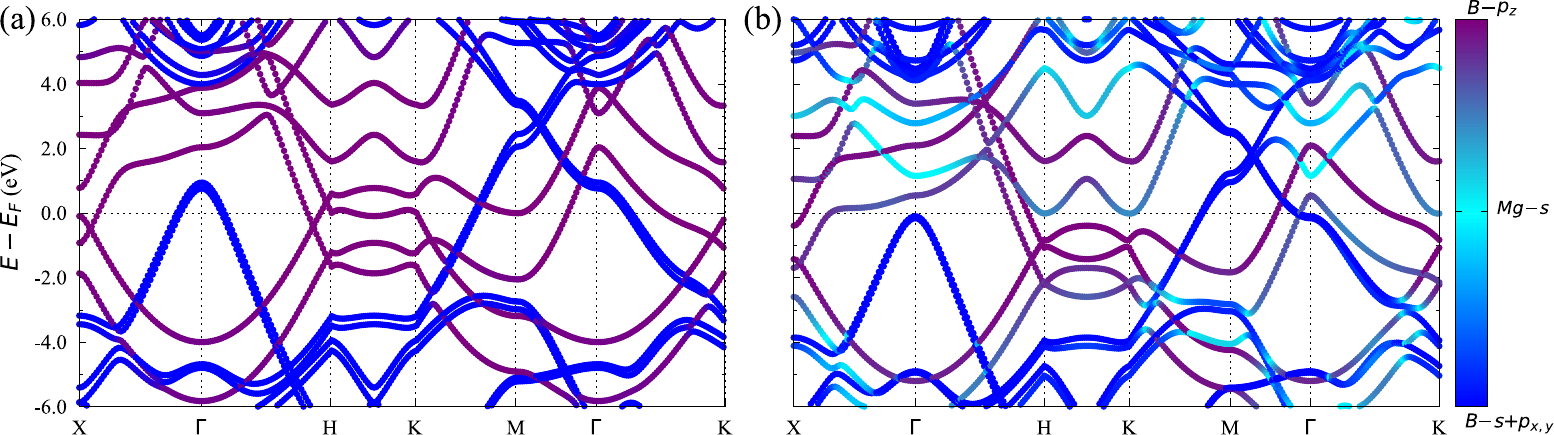}
\caption{\label{fig:pbands_1} Orbital-resolved bands of (a) pristine and (b) Mg-intercalated $\chi_{3}$ bilayer borophene. The blue, purple, and cyan dots represent the B-$s+p_{x,y}$, B-$p_{z}$, and Mg-$s$ contribution, respectively.}
\end{figure*}
\begin{figure*}[!htp]
    \includegraphics[width=\textwidth]{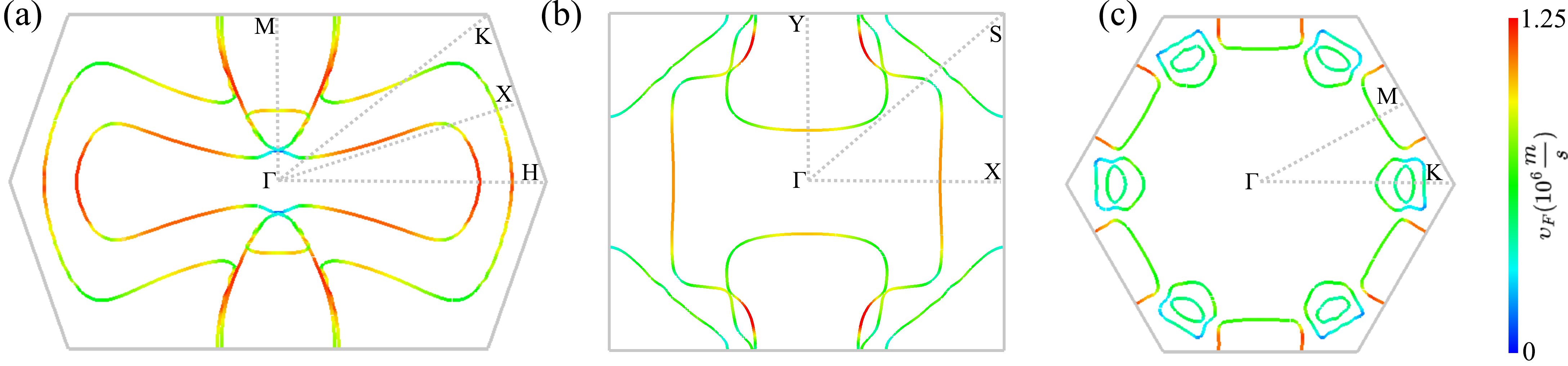}
\caption{\label{fig:fermi} Fermi surface (colored according to the Fermi velocities) for (a) Mg-intercalated $\chi_{3}$, (b) Be-intercalated $\delta_{4}$, and (c) Ca-intercalated kagome bilayer borophene.}
\end{figure*}
\subsubsection{\texorpdfstring{$\chi_{3}$}{chi3} phase of bilayer borophene \texorpdfstring{$(\eta=1/5)$}{eta_15}}
Compared to the $\beta_{12}$ phase with predominant contributions from out-of-plane orbitals close to $E_F$, the states near $E_F$ in the $\chi_{3}$ phase have both $\sigma$ and out-of-plane orbital character, as shown in Fig.~\ref{fig:pbands_1} for pristine and Mg-intercalated $\chi_3$ bilayers. To quantify this effect, we may consider the case of Mg intercalation, in view of its high total DOS at $E_F$ (1.71~states/eV) and its elevated $T_c$ of 17 K. We obtained contributions to the DOS at $E_{F}$ of 1.00~states/eV from B-$p_{z}$ (corresponding to out-of-plane states), 0.65~states/eV from B-$s+p_{x,y}$ orbitals ($\sigma$), and an additional 0.06~states/eV from the Mg $s$ orbital, as also depicted in Fig.~\ref{fig:pbands_1}(b). This mixture of $\sigma$ and out-of-plane states results in a plethora of intersecting Fermi contours with broadly varying Fermi velocities for the Mg intercalated $\chi_3$ bilayer compound, shown in Fig.~\ref{fig:fermi}(a). 

\begin{figure*}[!htp]
    \includegraphics[width=\textwidth]{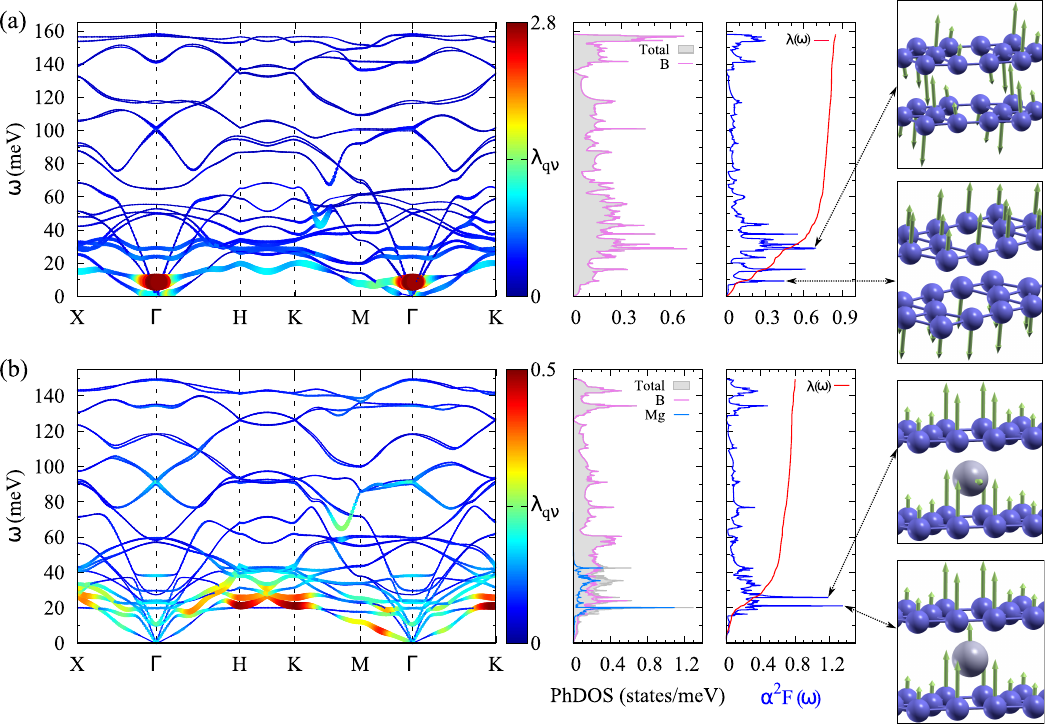}
\caption{\label{fig:phonons_1} Phonon band structure, atom-resolved phonon DOS, and the isotropic Eliashberg spectral function $\alpha^{2}F$ with modes corresponding to its peaks, of (a) pristine, and (b) Mg-intercalated $\chi_{3}$ bilayer borophene. The phonon branch ($\nu$) and $q$-resolved electron-phonon coupling $\lambda_{q\nu}$ is indicated by colors, as well as by point sizes that are proportional to $\lambda$.}
\end{figure*}

The analysis of the vibrational properties reveals that the $\chi_{3}$ phase is the only one of the investigated bilayer borophene compounds that is dynamically stable in pristine AA-stacked form. As depicted in Fig.~\ref{fig:phonons_1}, the predominant contributions to the total EPC for both pristine and Mg intercalated $\chi_3$ bilayers are situated within the frequency region below $\sim$50~meV. In the pristine case [Fig.~\ref{fig:phonons_1}(a)], the Eliashberg function features several prominent peaks in this range. The lowest two peaks correspond to an optical phonon branch ($A_{1}$ representation within the $C_{2V}$ point group), for which the EPC has maximum value of $\sim$2.8 at $\Gamma$. This mode at $\sim$9~meV consists of layer-breathing motion of the two boron sheets, as depicted in Fig.~\ref{fig:phonons_1}(a). The overall highest peak in the Eliashberg function of pristine bilayer $\chi_3$ (at $\sim$30~meV) originates from the optical $B_{2}$ phonon mode. At $\Gamma$, this mode consists of out-of-plane and out-of-phase movement of the B atoms -- as shown in Fig.~\ref{fig:phonons_1}(a) -- which are both infrared and Raman active (I+R). 

For the case of Mg intercalation [Fig.~\ref{fig:phonons_1}(b)], the EPC is notably strong along the H-K segment. The Eliashberg spectral function exhibits two distinct peaks, around 21~meV and 26~meV. The first one is related to the $A_{u}$ acoustic phonon mode, corresponding to out-of-plane motion of the Mg and B atoms, as depicted in Fig.~\ref{fig:phonons_1}(b). The second one is related to the $A_{g}$ optical phonon mode. With interesting multi-faceted Fermi contours, shown in Fig.~\ref{fig:fermi}(a), a total EPC of 0.8, and resulting $T_{c}$ of 17.3~K, Mg is the best candidate among all considered intercalants for superconductivity in the $\chi_3$ phase of bilayer borophene. 

The intercalated $\chi_3$ structures were also found to be dynamically stable with intercalants Li and Na. Upon intercalation with Li [see SM, Figs.\ S38(b) and (c)~\cite{noteSI}], the EPC within the $A_{1}$ phonon mode, at $\Gamma$, reduces to $\sim$0.6. The main two peaks appear at 29 and 33 meV, and stem from out-of-plane movement of the B atoms, and in-plane movement of Li atom, respectively. In the case of Na, there are also two main peaks in the Eliashberg function. The highest peak at 33 meV corresponds to the $A_{g}$ phonon mode at the X point, and again corresponds to out-of-plane movement of B atoms and in-plane movement of Na atom. Comparing the pristine form and the Na intercalated case, we observe a stronger contribution from high-frequency optical phonon modes to the EPC, along with an increase in the total electronic DOS and the logarithmically averaged phonon frequency ($\omega_{log}$). However, the total EPC is depleted, effecting a reduction of $T_c$ from 12.9~K to 9.1~K. 

\begin{figure*}[!htp]
    \includegraphics[width=\textwidth]{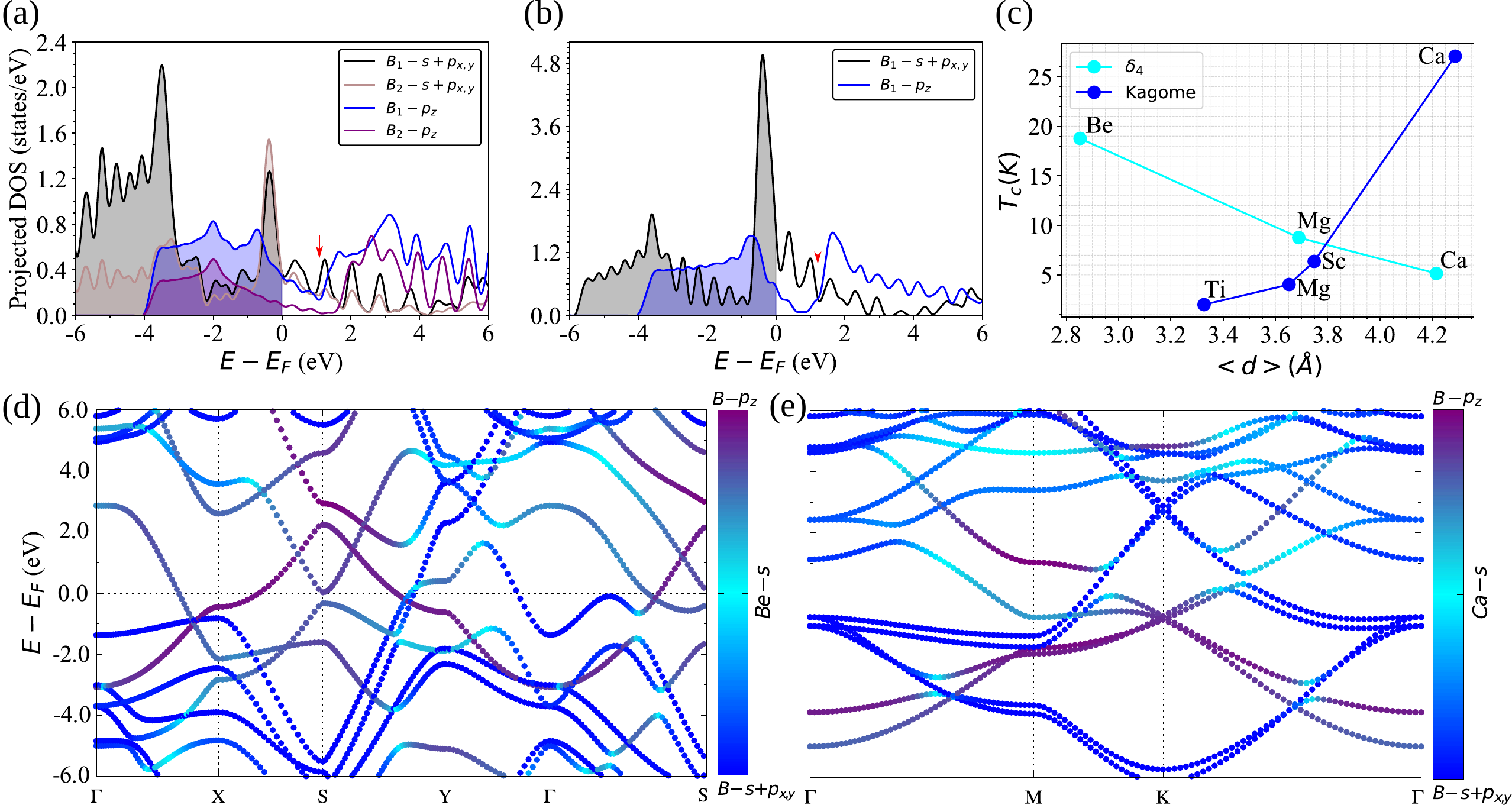}
\caption{\label{fig:pdos_tc} The projected density of states (PDOS) on $s+p_{x,y}$ and $p_{z}$ orbitals of pristine (a) $\delta_{4}$ and (b) kagome bilayer borophene. (c) The relation between the critical temperature and the average distance between the two borophene sheets in the two phases. The red arrows indicate the required shift of Fermi level in order to stabilize these systems. Additionaly, orbital-resolved bands of (d) Be-intercalated $\delta_{4}$ and (e) Ca-intercalated kagome bilayer borophene are also shown. The blue, purple, and cyan dots represent the B-$s$ + $p_{x,y}$, B-${p_z}$, and X-$s$ contribution, respectively.}
\end{figure*}

\begin{figure*}[!htp]
    \includegraphics[width=\textwidth]{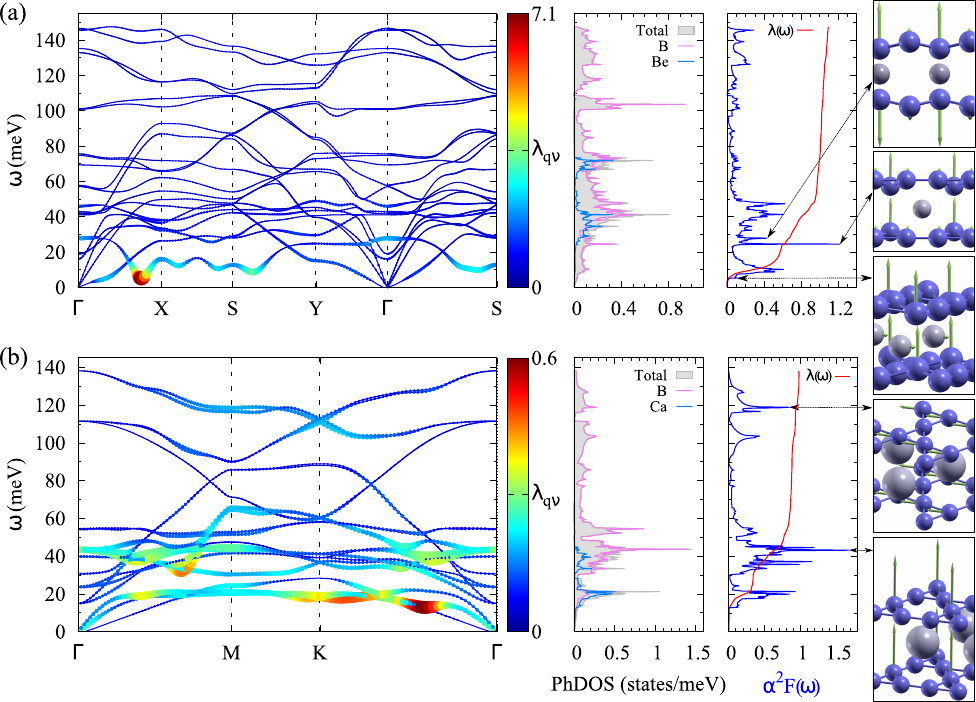}
\caption{\label{fig:phonons_2} Phonon band structure, atom-resolved phonon DOS, and isotropic Eliashberg spectral function $\alpha^{2}F$, together with modes corresponding to its peaks, of (a) Be-intercalated $\delta_{4}$, and (b) Ca-intercalated kagome bilayer borophene. The phonon branch ($\nu$) and $q$-resolved electron-phonon coupling $\lambda_{q\nu}$ is indicated by colors, as well by point sizes that are proportional to $\lambda$.}
\end{figure*}
\subsubsection{\texorpdfstring{$\delta_{4}$}{delta4} and kagome phase of bilayer borophene \texorpdfstring{$(\eta=1/4)$}{eta_14}}

Last but not least, we proceed to structures characterized by $\eta=1/4$. In contrast to the $\beta_{12}$ and $\chi_{3}$ phases, the $\sigma$ states ($s+p_{x,y}$) of $\delta_{4}$ and kagome phase are less occupied, therefore contribute significantly more to the DOS at $E_{F}$ as shown in Fig.~\ref{fig:pdos_tc}. Due to this electron deficiency the out-of-plane ($p_z$) states are partially filled, which renders these compounds dynamically unstable in their pristine form. Furthermore, such partially occupied $p_z$ states lead to weak interlayer interaction, resulting in larger interlayer distances (see Table~\ref{tab_a}). As can be observed in Fig.~\ref{fig:pdos_tc}(a) and (b), due to the partially filled $p_z$ orbitals, there is a distinct gap between $\pi$ and $\pi^{*}$ states above $E_{F}$. Moreover, for intercalated kagome bilayer borophene, we found van Hove singularities (saddle points at the M point), and high-order van Hove singularities (at the K point), intersecting or being near to the Fermi level, stemming from hybridized B-$p_z$ and intercalant $s$-state.  

Thus, to stabilize the $\delta_{4}$ and kagome phases, intercalants able to transfer sufficient electronic charge to the borophene layers are required. For example, by shifting the Fermi level by $\sim$ 1.2 eV, a good electronic balance between the $\sigma$ and $\pi^{*}$ states can be achieved, which is crucial for dynamical stability of the material. However, we found that the electron transfer from alkali metal intercalants is not sufficient to stabilize the borophene sheets (donating at most $\sim 0.87~e$ per intercalant). In contrast, dynamically stable structures were obtained by intercalation with alkaline-earth and transition elements, where the electron transfer is significantly larger (donating at least $\sim 1.20~e$ per intercalant). 

Despite the equal number of B atoms in their unit cells, the different crystal symmetries of the $\delta_{4}$ and kagome structures result in distinct local structural and electronic properties. This becomes particularly noticeable upon intercalation. For the $\delta_{4}$ phase we found significant buckling of borophene layers due to stronger bonding of alkaline-earth and transition metal intercalants with B$_{2}$ compared with B$_{1}$ atoms. On the other hand, as the kagome phase consists of only one type of atom, there is no buckling taking place. Therefore, the $\delta_4$ case is characterized by a nonuniform charge transfer from the intercalant to the boron atoms, while in the kagome phase the transfer is uniform and bond strengths are equal [see Electron Localization Function (ELF) plots in the SM, Figs. S24--S31~\cite{noteSI}]. This yields opposite trends in their $T_{c}$ values as a function of the (average) distances between the borophene layers, as shown in Fig.~\ref{fig:pdos_tc}(c).

Another characteristic separating the two phases concerns the relation between the $T_{c}$ and the DOS at $E_{F}$, with a clear correlation in the kagome case, which is absent for the $\delta_{4}$ phase. This is clearly seen for the Ca-intercalated kagome structure, with the highest DOS at $E_{F}$ among the $\eta=1/4$ structures -- 1.71~states/eV, contributed mainly by B-$p_{z}$ and B-$s+p_{x,y}$ orbitals -- and the resulting overall highest $T_{c}$ of 27~K. 

Let us now delve deeper into the superconducting properties of the Be-intercalated $\delta_{4}$ phase. This structure possesses an interesting Fermi surface, shown in Fig.~\ref{fig:fermi}(b), consisting of several straight parallel segments with strong nesting properties, in particular along $\Gamma-X$. At first glance, there is an unexpected highly softened phonon branch ($A_{1}$ representation within the $C_{2V}$ point group), with several dips that make a significant contribution to the Eliasberg function, as shown in Fig.~\ref{fig:phonons_2}(a). Thus, the first peak at $\sim$5~meV which possesses a very strong EPC value of 7.1, and the overall highest peak at $\sim$24~meV, both stem from the same mode. This mode corresponds to in-phase and out-of-plane movement of B$_{2}$ atoms together with small in-plane displacements of the Be atom. This behavior is caused by the strong interaction between the borophene layers, mediated by Be which forms a strong bond with the B$_{2}$ atoms. Additionally, this interaction causes a shift of the layer-breathing mode at $\Gamma$ to higher frequencies ($\sim$28~meV) when compared with the pristine $\chi_{3}$ phase [cf.\ Fig.~\ref{fig:phonons_1}(a)]. Furthemore, due to the different symmetry, this mode belongs to the $A_{g}$ representation (within the $D_{2h}$ point group), and it is Raman active. Similarly, in the cases of Mg and Ca intercalated $\delta_4$ bilayers, the highest EPC values are also stemming from softened phonon modes [see SM, Fig.\ S40~\cite{noteSI}]. In the case of Ca, these low-frequency peaks contribute less to the Eliashberg function, resulting in a lower $T_{c}$ of 5.2~K compared to the 8.8~K of the Mg case. 

Considering the superconducting properties of the Ca-intercalated kagome phase, one should notice its scarcely intersecting Fermi contours [see Fig.~\ref{fig:fermi}(c)] and overall highest $T_{c}$ of 27.1~K. As shown in Fig.~\ref{fig:phonons_2}(b), the maximal EPC value of 0.6 corresponds to the first peak in the Eliashberg function, related to the $A_{g}$ phonon mode (within point group $C_{s}$), entailing rotation of B atoms together with in-plane movement of Ca atoms. The highest peak corresponds to the flat segment in the phonon dispersion around M ($\sim$43~meV), comprising out-of-plane motion of the B atoms. At higher frequencies ($\sim$119~meV) we find another peak related to optical phonon mode of in-plane B atoms movement. The total EPC for Ca is high ($\lambda=0.98$), establishing it as the best candidate among all of the intercalants for the kagome phase of bilayer borophene. 

Upon intercalation of the kagome bilayers with transition metal elements, the total EPC reaches 0.61 for Sc, which is significantly higher than 0.43 from Ti, directly reflecting on their respective $T_{c}$'s of 6.4 and 2.0~K. 

In summary, similarly as for $\beta_{12}$ and $\chi_3$ phases, the highest obtained $T_{c}$ values in both $\delta_4$ and kagome structures were realized by intercalation with alkaline-earth elements -- Be in $\delta_{4}$ and Ca in the kagome phase. 

\section{Anisotropic Migdal-Eliashberg calculations}

\begin{figure}[!htp]
    \includegraphics[width=\linewidth]{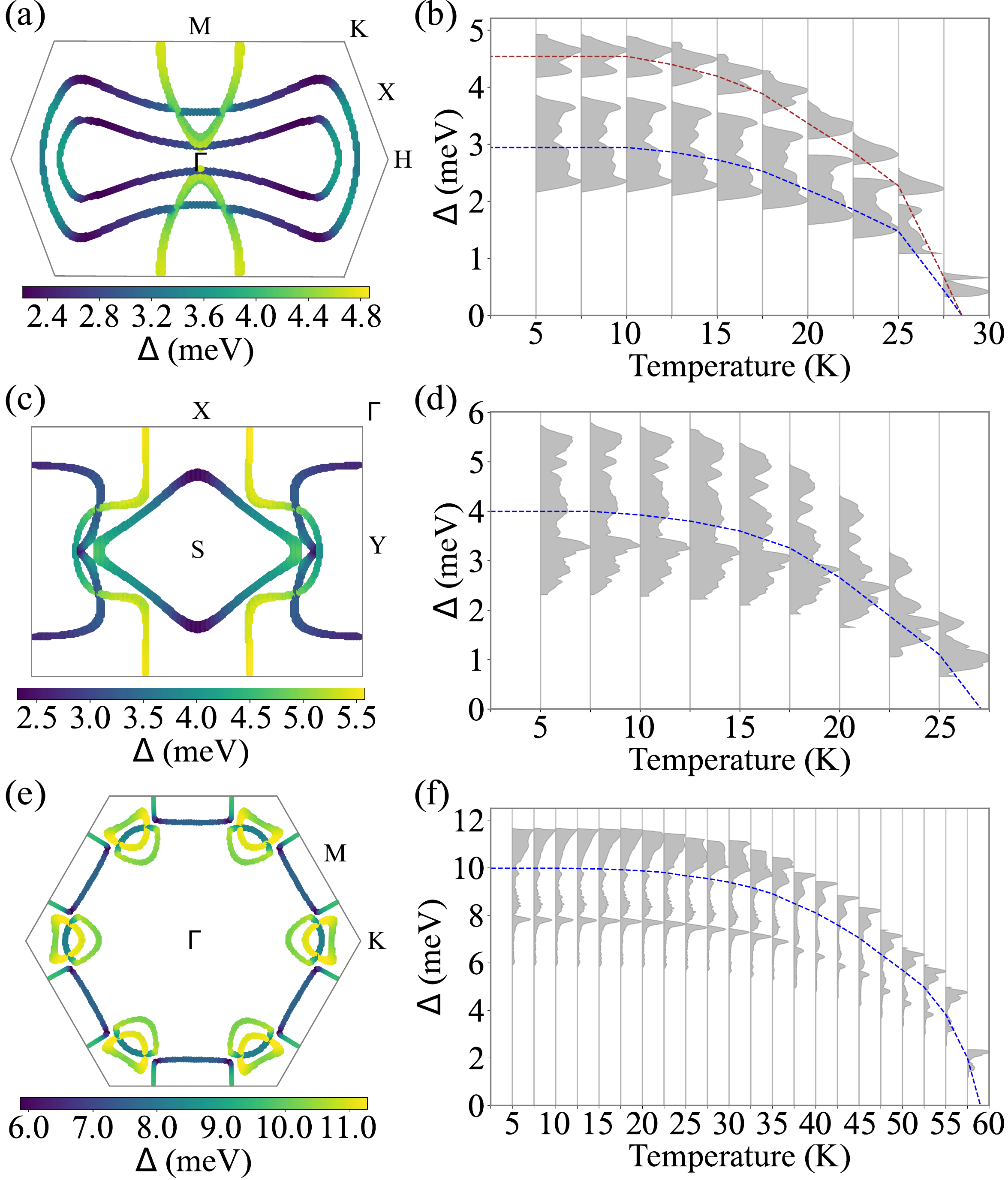}
\caption{\label{fig:anisotropic} Momentum-dependent superconducting gap on the Fermi surface at 5~K, and evolution of the superconducting gap distribution with temperature, for (a,b) Mg in $\chi_{3}$, (c,d) Be in $\delta_{4}$, and (e,f) Ca in kagome bilayer borophene, respectively.}
\end{figure}

Taking into account that the Fermi surfaces in Fig.~\ref{fig:fermi} exhibit strong anisotropy, we have subsequently performed fully anisotropic Migdal-Eliashberg calculations to completely characterize the superconducting properties of the prime candidates. The results are summarized in Fig.~\ref{fig:anisotropic}.

The Fermi surface of Mg-intercalated $\chi_3$ bilayer borophene, shown in Fig.~\ref{fig:anisotropic}(a), consists of several intersecting Fermi contours. The ellipsoidal B-$\sigma$ Fermi sheets located around $\Gamma$-M path host the strongest gap values within the anisotropic gap spectrum $\Delta (\mathbf{k})$. At low temperatures, the  $\Delta$ values stemming from B-$\sigma$ states reach 4.9~meV, while the gap of the B-$p_z$ and Mg-$s$ states is highly anisotropic, ranging from 2.2 to 3.9~meV. This results in distinctly two-gap superconductivity. By solving the anisotropic Migdal-Eliashberg equations for a range of temperatures, we found that the anisotropic $T_c$ of this compound is 28.5~K.

For Be-intercalated $\delta_4$ bilayer borophene, there is hybridization of different states, resulting in overlapping gaps on the Fermi surface. Fig.~\ref{fig:anisotropic}(c) shows this anisotropic single-gap nature. The $\Delta$ values reach 5.6~meV at low temperatures, at the Fermi contour mainly stemming from strongly hybridized B-$p_z$ and Be-$s$ states -- specifically along $\Gamma$-X segment where strong nesting is present. The Fermi contour around the S point mainly carries B-$p_z$ character, and has a much smaller minor contribution to $\Delta$, and similarly for the Fermi contour along the $\Gamma$-Y segment with dominant B-$\sigma$ character. The resulting anisotropic $T_c$ is 27~K, as shown in Fig.~\ref{fig:anisotropic}(d).

Finally, we arrive at the best candidate for superconductivity among the studied structures, namely Ca-intercalated kagome bilayer borophene. This compound harbors a single, strongly anisotropic gap. The gap reaches a maximum value of 11.4~meV at low temperatures [see Fig.~\ref{fig:anisotropic}(e)]. The highest contribution mainly stems from B-$\sigma$ states positioned along $\Gamma$-K. The weaker $\Delta$ contributions stem from the Fermi sheet situated along $\Gamma$-M, corresponding to hybridized B-$p_z$ and Ca-$s$ states. As shown in Fig.~\ref{fig:anisotropic}(f), the anisotropic $T_c$ is 58~K, establishing Ca-intercalated kagome bilayer borophene as the best candidate for robust superconductivity at elevated temperatures among all the compounds considered in this work.


\section{Conclusions}

In summary, we have performed an extensive first-principles exploration of the boron-based 2D superconductors, specifically metal-intercalated bilayer borophene phases, considering four bilayer borophene phases ($\beta_{12}$, $\chi_3$, $\delta_4$ and kagome) and nine possible intercalants (Li, Na, K, Be, Mg, Ca, Sc, Ti, V). For a comparative analysis of the emergent superconductivity in these materials, we have employed the isotropic Eliashberg formalism combined with \textit{ab initio} calculations of the electronic and vibrational properties. In total, we identified seventeen possible superconductors, with diverse electronic and vibrational properties, and isotropic $T_{c}$ values ranging up to 27.1~K. 

We found that among all considered structures, only in the case of intercalated $\beta_{12}$ phase the $\sigma$ states are almost fully occupied, with a clear gap separating them from the $\sigma^{*}$ states. Therefore, the contribution of these states at $E_{F}$ is low, and the metallic character is mainly provided by the $p_{z}$ orbitals. This strongly limits the isotropic $T_{c}$ to low values of $0.1-1$~K. 

On the other hand, in the $\chi_{3}$ phase ($\eta=1/5$), the metallic properties are governed by both $\sigma$ and out-of-plane ($p_{z}$) states, mainly due to the intrinsically larger electronic deficit in the $\chi_3$ monolayers. The contribution of $\sigma$ states at $E_{F}$ is larger in this case, and as a result the isotropic $T_{c}$ values were found to be higher compared to the $\beta_{12}$ case, ranging from 9.1 to 17.3~K. 

Proceeding to phases with $\eta=1/4$, namely $\delta_4$ and kagome, the electron deficit becomes even larger, resulting in a distinct gap between the $\pi$ and $\pi^{*}$ states. Only intercalants from the group of the alkaline-earth and transition metal elements can provide sufficient electronic transfer to the borophene layers to stabilize them. The contribution of the $\sigma$ states is also high, hence due to this combined effect the isotropic $T_{c}$ reaches values up to 27.1~K for the Ca-intercalated kagome bilayer. 

Our results therefore suggest that the combination of the stabilization of the out-of-plane states and the contribution of the $\sigma$ states at $E_{F}$ (which play an important role in the coupling with in-plane vibrational modes) yields higher isotropic $T_{c}$ values. We also found that, generally, a larger contribution from the B-$s+p_{x,y}$ orbitals to the DOS at $E_{F}$ limits the EPC strength from phonon modes with frequencies above $\sim$50~meV (as visible in the Eliashberg function), and vice versa. The exception to this rule are intercalants from the group of transition metals, where $d$ states play a more important role. Furthermore, higher $\eta$ values align with higher isotropic $T_{c}$ values -- a discernible trend in our study, that can be connected to MBene structures with $\eta=1/3$, for which even higher isotropic $T_{c}$ values (up to 36 K) have been reported~\cite{cahex}.

After having revealed distinct trends in the superconducting properties of various intercalated borophene bilayers, we have delved deeper into the analysis of the electronic origin of superconductivity by considering anisotropic and multiband effects through anisotropic Eliashberg calculations for selected compounds. We have thereby revealed the high importance of the anisotropy of the Fermi surface, and that it contributes to a two-gap behavior of Mg-intercalated $\chi_3$ bilayer borophene as well as to elevated $T_c$ values in all compounds, of up to 58~K in the case of the Ca-intercalated kagome bilayer borophene. An interesting step for further exploration would be the inclusion of anharmonic effects, expected to play an important role in the case of the lighter elements~\cite{anh}. Moreover, in contrast to their monolayer counterparts, the functionalized bilayer borophenes studied here are expected to be more chemically stable and more resilient to contamination and deterioration such as due to oxidation. Considering that several phases in this work have exhibited robust superconductivity and elevated $T_{c}$ values, in particular the ones intercalated with alkaline-earth metals, our results position the intercalated bilayer borophenes among the prime candidates for further development of boron-based two-dimensional superconducting technologies.

\section*{Acknowledgments}
This work was supported by the Research Foundation-Flanders (FWO). B.N.Š. acknowledges support from the Montenegrin Ministry of Science and the Special Research Fund of the University of Antwerp (BOF-UA) -- No.\ 542300011. J.B. acknowledges support as a Senior Postdoctoral Fellow of Research Foundation-Flanders (FWO) under Fellowship No.\ 12ZZ323N. The computational resources and services were provided by the VSC (Flemish Supercomputer Center), funded by the FWO and the Flemish Government -- department EWI. The collaborative effort in this work was fostered by EU-COST actions NANOCOHYBRI (Grant No.\ CA16218) and SUPERQUMAP (Grant No.\ CA21144), and Erasmus KA107 - Mobility program between the University of Montenegro and the University of Belgrade.

\nocite{*}

\bibliography{bs}

\end{document}